\documentclass[12pt,conference]{IEEEtran}

\usepackage{subcaption}
\usepackage{ifpdf}
\ifCLASSINFOpdf
\usepackage{graphicx}
\usepackage{grffile}
\usepackage{epstopdf}
\usepackage{grffile}
\usepackage[export]{adjustbox}
\else
\fi
\usepackage[cmex10]{amsmath}
\usepackage{array}
\usepackage{booktabs}
\usepackage{lipsum}
\usepackage{multicol}
\usepackage{mathtools}
\usepackage{bm}
\usepackage{apacite}

\usepackage{abstract}
\begin{document}
\title{Revisit of the Eigenfilter Method for the Design of FIR Filters and Wideband Beamformers}
\author{\IEEEauthorblockN{Ahsan Raza and Wei Liu}\\
\IEEEauthorblockA{Communications Research Group}
{Department of Electronic and Electrical Engineering}\\
{University of Sheffield, S1 3JD, U.K.}\\
  { \tt \{smajafri1, w.liu\}@sheffield.ac.uk }
  }
\twocolumn[
\begin{@twocolumnfalse}
\maketitle
%%%%%%%%%%                                                                ABSTRACT                                                    %%%%%%%%%%%%%%
\begin{abstract}
The least squares based eigenfilter method has been applied to the design of both finite impulse response (FIR) filters and wideband beamformers successfully. It involves calculating the resultant filter coefficients as the eigenvector of an appropriate Hermitian matrix, and offers lower complexity and less computation time with better numerical stability as compared to the standard least squares method. In this paper, we revisit the method and critically analyze the eigenfilter method by revealing a serious performance issue in the passband of the designed FIR filter and the mainlobe of the wideband beamformer,  which occurs due to a formulation problem. A solution is then proposed to mitigate this issue by imposing an additional constraint to control the response at the passband/mainlode, and design examples for both FIR filters and wideband beamformers are provided to demonstrate the effectiveness of the proposed method.\footnotemark
\end{abstract}
\vspace*{-\baselineskip}
% no keywords
\begin{IEEEkeywords}
\textbf{Keywords:} least squares, eigenfilter, filter design, wideband beamformer, constrained design.
\end{IEEEkeywords}
 \end{@twocolumnfalse}
]
\footnotetext{This is an expanded work of our conference publication \cite{ahsan1}}
\IEEEpeerreviewmaketitle

%%%%%%%%% %%%                              SECTION 1     INTRODUCTION                       %%%%%%%%%%%%%

\section{\textbf{Introduction}}
% no \IEEEPARstart

FIR filters and wideband beamformers have numerous applications ranging from Sonar, Radar, audio processing, ultrasound imaging, radio astronomy, earthquake prediction, medical diagnosis, to communications, etc \cite{vantrees02a,liu10g}. Many optimization methods have been employed in the past to design FIR filters and wideband beamformers with required specifications. General convex optimization is one of the techniques that has been extensively explored from this perspective ~\cite{elkeyi05a,ahsan01,duan08a,liu11d} with the inherent drawback of long computation time required to reach a feasible solution. 

Although it can be considered as a special case of the convex optimization approach, least squares based design has been adopted as a simple but effective solution to both design problems, which minimizes the mean squared error between the desired and designed responses ~\cite{liu10g,liu11f,doclo03b}. The solution of the standard least squares cost function involves matrix inversion to obtain the required weight vector. Since matrix inversion poses numerical instability with long filters \cite{tkacenko03a}, another method was proposed based on the least squares approach by performing eigenvector decomposition of a cost function to extract the required weight vector in the form of an eigenvector. This method is called eigenfilter design and has been explored for designing both filters and beamformers ~\cite{vaidyanathan87a,nguyen93a,pei01a,chen02aa,doclo02a,liu11c}. Moreover, the design of linear-phase FIR Hilbert transformers and arbitrary order digital differentiators were considered by Pei and Shyu ~\cite{pei88a,pei89a}, who also investigated the design of nonlinear-phase filters with arbitrary complex-valued coefficients~\cite{pei92a,pei93a}. Two-dimensional (2-D) extension to the eigenfilter method was proposed by Nashashibi and Charalambous \cite{nashashibi88a}, and later considered by Pei ~\cite{pei90a,pei93b}. Eigenfilters have also been used to design infinite impulse response (IIR) and all-pass filters ~\cite{laakso93a,shyu92a}.

In this work, we revisit the eigenfilter method for designing FIR filters and wideband beamformers and reveal a serious performance issue in the passband of the designed FIR filters and the mainlobe of the designed wideband beamformers  in the light of an inherent design formulation flaw. An overall critical analysis of the performance of this approach is presented with the suggested modification for tackling this issue. In particular, an additional constraint is imposed at the passband/mainlode of the system to control the resultant responses.

This paper is organized as follows. The eigenfilter based design formulation for FIR filters and wideband beamformers along with the critical analysis is presented in Section \ref{sec:lsdesign}. The proposed solution to the highlighted problem is given in Section \ref{sec:proposed_sol}. Design examples for different types of FIR filters and wideband beamformers affected by the problem are provided in Section \ref{sec:design_examples} followed by results using the proposed solution. Conclusions are drawn in Section \ref{sec:conclusion}.

%%%%%%%%%%%%                   SECTION 2    EIGENFILTER BASED DESIGN OF FIR FILTERS AND WIDEBAND BEAMFORMER        %%%%%%%%%%%%%

\section{\textbf{Least squares based design and critical analysis}}
\label{sec:lsdesign}

%%%%%%%%%   SUBSECTION 2A EIGENFILTER FIR FILTER DESIGN %%%%%%%%%%%%

\subsection{\textbf{FIR filter design}}
\label{sec:lsdesign_filter}
 Consider an $N-$tap FIR filter. Its frequency response $W(e^{j\omega})$ is given by
\begin{equation}
 W(e^{j\omega})=\sum_{n=0}^{N-1}w_n e^{-jn\omega} \;,
 \end{equation}
where $w_n$ is the $n-$th tap/coefficient of the filter. In vector form, it can be expressed as
\begin{equation}
\label{eq:weight}
W(e^{j\omega})= \textbf{w}^H\textbf{c}(\omega)\;,
\end{equation}
where  $\textbf{w}$ is the $N\times 1$ weight vector holding the coefficients   $w_n$, $n=0, 1, \dots, N-1$, and
\begin{equation}
 \textbf{c}(\omega)={[1, e^{-j\omega}, \cdots, e^{-j(N-1)\omega}]}^T\;.
\label{eq:filter_vector}
 \end{equation}

Now consider designing a lowpass filter as an example. The desired response $D(\omega)$ is given by
\begin{equation}
D(\omega) = \left\{ \,
\begin{IEEEeqnarraybox}[][c]{l?s}
\IEEEstrut
e^{-j\omega\frac{N-1}{2}}, & $0 \leq \omega \leq \omega_p$ \\
0, & $\omega_s \leq \omega \leq \pi$ %
%\textrm {Don't care} , & $\omega_p \leq \omega \leq \omega_s$
\IEEEstrut
\end{IEEEeqnarraybox}
\right.
\label{eq:example_left_right1}
\end{equation}
where $e^{-j\omega\frac{N-1}{2}}$ represents the desired linear phase at the passband with a delay of $\frac{N - 1}{2}$ samples along with the desired stopband response equal to zero.

The design process involves formulating the cost function in the standard eigenfilter form, based on the Rayleigh-Ritz principle which states that for any Hermitian matrix $\textbf{R}$, its Rayleigh-Ritz ratio is given by
\begin{equation}
\label{eq:rayleigh}
\frac{\textbf{w}^H\textbf{R}\textbf{w}}{\textbf{w}^H\textbf{w}}\;.
\end{equation}
This ratio reaches its maximum/minimum when $\textbf{w}$ is the eigenvector corresponding to the maximum/minimum eigenvalue of  $\textbf{R}$. The maximum and minimum values of this ratio are respectively the maximum and minimum eigenvalues. For FIR filter design, a reference frequency point was introduced by Nguyen in the passband region of the cost function to help represent it into the quadratic form as desired by \eqref{eq:rayleigh}~\cite{nguyen93a}. The cost function with the reference frequency point incorporated is given as

\begin{equation}
E=\frac{1}{\pi}\int_{\omega}v(\omega) \left| \frac{D(\omega)}{D(\omega_r)}W(e^{j\omega_r})-W(e^{j\omega})\right|^2d\omega \end{equation}
where  $v(\omega)$ is the weighting function and $D(\omega_r)$ and $W(e^{j\omega_r})$ represent the desired and designed responses at reference frequency,  respectively. This expression can also be written as
\begin{equation}
\label{eq:expand}
\begin{aligned}
E=\frac{1}{\pi}\int_{\omega}v(\omega) \left(\frac{D(\omega)}{D(\omega_r)}W(e^{j\omega_r})-W(e^{j\omega})\right)\\
\left(\frac{D(\omega)}{D(\omega_r)}W(e^{j\omega_r})-W(e^{j\omega})\right)^H d\omega
\end{aligned}
\end{equation}
For stopband, the desired response  $D(\omega)=0$. Substituting this value into the expression above, we have
\begin{equation}
\label{eq:stopband}
E_s=\frac{1}{\pi}\int_{\omega_s}^{\pi}v(\omega) W(e^{j\omega}) W(e^{j\omega})^H d\omega
\end{equation}
Substituting the expression in \eqref{eq:weight} into \eqref{eq:stopband}, the expression further simplifies to
\begin{equation}
\label{eq:stopband_exp1}
E_s=\frac{1}{\pi}\int_{\omega_s}^{\pi} v(\omega)\textbf{w}^H\textbf{c}(\omega)\textbf{c}(\omega)^H\textbf{w}d\omega
\end{equation}
Then we can express \eqref{eq:stopband_exp1} as
\begin{equation}
E_s=\textbf{w}^H\textbf{P}_s\textbf{w}
\end{equation}
where $\textbf{P}_s$ is a symmetric, positive definite matrix of order $N$ x $N$ given by
\begin{equation}
\label{eq:stopband_exp}
\textbf{P}_s = \frac{1}{\pi}\int_{\omega_s}^{\pi} v(\omega)\textbf{c}(\omega)\textbf{c}(\omega)^Hd\omega
\end{equation}

The passband cost function is derived by incorporating the desired passband response $D(\omega)=e^{-j\omega\frac{N-1}{2}}$  into \eqref{eq:expand}
\begin{equation}
\begin{aligned}
E_p=\frac{1}{\pi}\int_{0}^{\omega_p}v(\omega)\left(\frac{e^{-j\omega\frac{N-1}{2}}}{e^{-j\omega_r\frac{N-1}{2}}}W(e^{j\omega_r})-W(e^{j\omega}) \right) \\
\left(\frac{e^{-j\omega\frac{N-1}{2}}}{e^{-j\omega_r\frac{N-1}{2}}}W(e^{j\omega_r})-W(e^{j\omega})\right)^H d\omega
\end{aligned}
\end{equation}
After simplification, we have
\begin{equation}
\begin{aligned}
 E_p=\frac{1}{\pi}\int_{0}^{\omega_p} v(\omega) \textbf{w}^H  \left(e^{-j\frac{N-1}{2}(\omega-\omega_r)}  \textbf{c}(\omega_r)-\textbf{c}(\omega)\right) \\
 \left(e^{-j\frac{N-1}{2}(\omega-\omega_r)} \textbf{c}(\omega_r)-\textbf{c}(\omega)\right)^H \textbf{w}  d\omega
\end{aligned}
\end{equation}
This expression can also be written as
\begin{equation}
E_p=\textbf{w}^H\textbf{P}_p\textbf{w}\;,
\end{equation}
where  $\textbf{P}_p$ is a symmetric, positive definite matrix of order $N$ x $N$ given by
\begin{equation}
\label{eq:passband_exp}
\begin{aligned}
 \textbf{P}_p=\frac{1}{\pi}\int_{0}^{\omega_p} v(\omega) \left(e^{-j\frac{N-1}{2}(\omega-\omega_r)}  \textbf{c}(\omega_r)-\textbf{c}(\omega) \right) \\
 \left(e^{-j\frac{N-1}{2}(\omega-\omega_r)} \textbf{c}(\omega_r)-\textbf{c}(\omega)\right)^H  d\omega
\end{aligned}
\end{equation}
The total cost function is a combination of the passband and stopband cost functions with a trade-off factor $\alpha$
\begin{equation}
E=\alpha E_p+(1-\alpha)E_s\;, \;\;\;\;\; 0\leq \alpha \leq 1\;,
\end{equation}
which can be transformed into
\begin{equation}
E=\textbf{w}^H\textbf{P}\textbf{w}\;,
\end{equation}
where
\begin{equation}
\label{eq:final}
\textbf{P}=\alpha \textbf{P}_p+(1-\alpha)\textbf{P}_s,  0\leq \alpha \leq 1\;.
\end{equation}

Combining \eqref{eq:stopband_exp} and \eqref{eq:passband_exp} in \eqref{eq:final} and taking the real part, we have
\begin{equation}
\label{eq:filter_exp}
\begin{aligned}
 \textbf{P}=\alpha \int_{0}^{\omega_p} \textbf{Re} [ \left(e^{-j\frac{N-1}{2}(\omega-\omega_r)}  \textbf{c}(\omega_r) - \textbf{c}(\omega)\right) \\
\left(e^{-j\frac{N-1}{2}(\omega-\omega_r)} \textbf{c}(\omega_r)-\textbf{c}(\omega)\right) ^H] d\omega\\
+(1-\alpha) \int_{\omega_s}^{\pi}\textbf{Re} [\textbf{c}(\omega)\textbf{c}(\omega)^H] d\omega
\end{aligned}
\end{equation}
The solution rests in finding the eigenvector $\textbf{w}$ corresponding to the minimum eigenvalue of $\textbf{P}$ which minimizes $E$. The norm constraint $\textbf{w}^H \textbf{w}=1$ is also incorporated to avoid trivial solution. The final expression of solution for the eigenfilter based FIR filter design problem is given by
\begin{equation}
\label{eq:filtereigen_sol}
\begin{aligned}
& \underset{\textbf{w}}{\text{Min}}
& & \frac{{\textbf{w}}^H\textbf{P}\textbf{w}} {\textbf{w}^H\textbf{w}}
\end{aligned}
\end{equation}
After investigating the designed filter's performance, it is found that although the design performs well for most of the cases with varying specifications for short filters, it produces ever increasingly inconsistent results as the number of filter taps increases for the same set of specifications. With those longer filters, the passband performance starts varying and switches from one case with flatness around near unity gain to another case with flatness achieved at almost zero magnitude. %In some cases, the passband attains such a low magnitude that the ratio between passband and stopband reduces to a very small value resulting in a non-feasible filter design.

This unstable performance can be attributed to the formulation in \eqref{eq:filter_exp} where the first part of the cost function measures the difference between the filter's response at the reference frequency $\omega_r$ and those at the other frequencies $\omega$ in the passband. The term $e^{-j\frac{N-1}{2}(\omega-\omega_r)}$ compensates for different phase shifts of the response at different frequencies. This expression minimizes the relative variation of the filter's response at different passband frequencies and ensures a flat passband response. However, there is no control over the absolute value of the filter's response in passband, allowing any type of flat passband response with arbitrary absolute magnitude leading to inconsistent design performance.

%%%%%%%%%%%%%%              SUBSECTION 2B WIDEBAND BEAMFORMER DESIGN                      %%%%%%%%%%%%%

\subsection{\textbf{Wideband beamformer design}}
\label{sec:lsdesign_wideband}
\begin{figure}
\centering
\includegraphics[width=.5\textwidth]{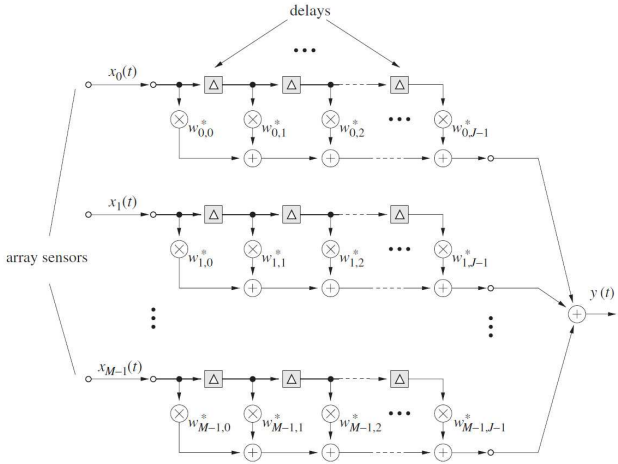}
\caption{A general structure for wideband beamforming.}
\label{fig:wideband}
\end{figure}
Consider a wideband beamformer with tapped delay lines (TDLs) or FIR filters shown in Figure \ref{fig:wideband}, where $J$ is the number of delay elements associated with each of the $M$ sensors. The wideband beamformer samples the propagating wave field in both space and time. Its response as a function of signal angular frequency $\omega$ and direction of arrival $\theta$ is given by \cite{liu10g}
\begin{equation}
\label{eq:wideband}
P(\omega,\theta)=\sum_{m=0}^{M-1}\sum_{k=0}^{J-1}w_{m,k}e^{-j\omega(\tau_m+kT_s)}\;,
\end{equation}
where $T_s$ is the delay between adjacent taps of the TDL and $\tau_m$ is the spatial propagation delay between the $m-th$ sensor and the reference sensor. We can also express \eqref{eq:wideband} as
\begin{equation}
\label{eq:wideband_firstexp}
P(\omega,\theta)=\textbf{w}^T\textbf{d}(\omega,\theta)\;,
\end{equation}
where $\textbf{w}$ is the coefficient vector
\begin{equation}
\textbf{w}={[w_{0,0}, \cdots w_{M-1,0}, \cdots w_{0,J-1}, \cdots, w_{M-1,J-1}]}^T
\end{equation}
and $\textbf{d}(\omega,\theta)$ is the $M$ x $J$ steering vector
\begin{equation}
\textbf{d}(\omega,\theta)=\textbf{d}_{Ts}(\omega)\otimes \textbf{d}_{\tau_m}(\omega,\theta)\;,
\end{equation}
where $\otimes$ denotes the Kronecker product. The terms $\textbf{d}_{Ts}(\omega)$ and  $\textbf{d}_{\tau_m}(\omega,\theta)$   are defined as
\begin{equation} 
\textbf{d}_{Ts}(\omega)= {[1, e^{-j\omega T_s}, \cdots, e^{-j(J-1)\omega T_s}]}^T
\end{equation}
\begin{equation} 
\textbf{d}_{\tau_m}(\omega,\theta)= {[e^{-j\omega \tau_0}, e^{-j\omega \tau_1}, \cdots, e^{-j\omega \tau_{M-1}}]}^T \;.
\end{equation}
For a uniform linear array (ULA) with an inter-element spacing $d$, and angle $\theta$ measured from the broadside, the spatial propagation delay $\tau_m$ is given by $\tau_m=m\tau_1=\frac{md\sin\theta}{c}$. With normalized angular frequency, $\Omega=\omega T_s$,  and $\mu=\frac{d}{cT_s}$, the steering vector is given by
\begin{equation} \textbf{d}(\Omega,\theta)=\textbf{d}_{T_s}(\Omega)\otimes \textbf{d}_{\tau_m}(\Omega,\theta)
\end{equation}
\begin{equation}
 \textbf{d}_{T_s}(\Omega)= {[1,e^{-j\Omega},\cdots,e^{-j(J-1)\Omega}]}^T
\end{equation}
\begin{equation}
 \textbf{d}_{\tau_m}(\Omega,\theta)= {[1,e^{-j\mu\Omega sin\theta},\cdots,e^{-j(M-1)\mu\Omega sin\theta}]}^T
 \end{equation}

 Now we have \eqref{eq:wideband_firstexp} as a function of $\Omega$ and $\theta$, given by
\begin{equation}
P(\Omega,\theta)=\textbf{w}^T\textbf{d}(\Omega,\theta)
\end{equation}

The desired response for the wideband beamformer is represented by $P_d(\Omega,\theta)$. Then, the eigenfilter based cost function can be expressed as
\begin{equation}
\begin{aligned}
J_{ef}(\textbf{w}) = \int_{\Omega_{pb}}\int_{\Theta}v(\Omega,\theta)\\
{\left|P(\Omega,\theta)- P(\Omega_r,\theta_r)\frac{P_d(\Omega,\theta)}{P_d(\Omega_r,\theta_r)} \right|}^2 d\Omega d\theta
\end{aligned}
\end{equation}
where $(\Omega_r,\theta_r)$ is the reference point. We can change this expression into
\begin{equation}
J_{ef}(\textbf{w}) = \textbf{w}^H \textbf{G}_{ef} \textbf{w}
\end{equation}
where
\begin{equation}
\label{eq:wideband_eigen}
\begin{aligned}
\textbf{G}_{ef}=\int_{\Omega_{pb}}\int_{\Theta}v(\Omega,\theta) \\
\left( \textbf {d}(\Omega,\theta)-\textbf{d}(\Omega_r,\theta_r)\frac{P_d(\Omega,\theta)}{P_d(\Omega_r,\theta_r)}\right) \\
{ \left( \textbf{d}(\Omega,\theta)-\textbf{d}(\Omega_r,\theta_r)\frac{P_d(\Omega,\theta)}{P_d(\Omega_r,\theta_r)}\right)}^H d\Omega d\theta
\end{aligned}
\end{equation}

Consider a typical design case with desired sidelobe response equal to zero and response at look direction $\theta_{0}$ given by $e^{-j \frac{J}{2}\Omega}$ equal to a pure delay; $\Omega_{r}$ and $\Omega_{pb}$ represent the reference frequency and passband frequency range, respectively, and $\alpha$ is the weighting factor for the mainlobe. The expression in \eqref{eq:wideband_eigen} is modified accordingly for real-valued beamformer coefficients and given by
\begin{equation}
\label{eq:wideband_finaleigen}
\begin{aligned}
\textbf{G}_{ef}=\alpha \int_{\Omega_{pb}} \textbf{Re} [\left (\textbf{d}(\Omega,\theta_0)- e^{-j\frac{J}{2}(\Omega-\Omega_r)} \textbf{d}(\Omega_r,\theta_r)\right) \\
\left(\textbf{d}(\Omega,\theta_0)- e^{-j\frac{J}{2}(\Omega-\Omega_r)} \textbf{d}(\Omega_r,\theta_r)\right) ^H] d\Omega\\
+(1-\alpha) \int_{\Omega_{pb}} \int_{\Theta_{sl}} \textbf{Re} [\textbf{d}(\Omega,\theta)\textbf{d}(\Omega,\theta)^H] d\Omega d\theta
\end{aligned}
\end{equation}
Then, the solution to the wideband beamformer design problem is given by
\begin{equation}
\begin{aligned}
& \underset{\textbf{w}}{\text{Min}}
& & \frac{\textbf{w}^H\textbf{G}_{ef}(\Omega,\theta)\textbf{w}}{{\textbf{w}^H\textbf{w}}}
\end{aligned}
\end{equation}

Similar to the FIR filter design case, testing of the designed wideband beamformer through the eigenfilter method showed an inconsistent design performance. The design performed well for some look directions, while attained a very poor response for other look directions.

This variable nature of look direction response for the same set of specifications can again be traced back to the design formulation in \eqref{eq:wideband_finaleigen}, where the first part of the expression calculates the difference between the beamformer response at reference point $(\Omega_r,\theta_r)$  and those at other frequencies in the look direction $\theta_0$ . The term $e^{-j\frac{J}{2}(\Omega-\Omega_r)}$ compensates for the different phase shifts experienced by the wideband signal at different frequencies. The formulation ensures minimzation of the relative error at the look direction for different frequencies, thus providing flat response at $\theta_0$. However, just like the FIR filter case, there is a lack of control for exact response in the look direction which can lead to design failure.

\section{\textbf{Proposed Solution with an Additional Constraint}}
\label{sec:proposed_sol}
%%%%%%%%%%%%%%%                               SECTION 3 PROPOSED SOLUTION                         %%%%%%%%%%%%%%%

As shown in our analysis of the eigenfilter design for both FIR filters and wideband beamformers in Section \ref{sec:lsdesign}, the key issue is its lack of control of the achieved response at the passband/look direction compared to the desired one in the formulation. To solve this problem, we add an additional constraint to the formulation to specify the required response explicitly at the reference point. Since the original formulation will minimize the variation of the achieved response in the passband/look direction, the explicit control of the response of the designed filter/beamformer at one reference point of the passband/look direction will guarantee the design reaches the desired response for the whole considered passband/look direction region with a minimum overall error.

 Now, constraining the reference frequency response to unity by adding a linear constraint to \eqref{eq:filtereigen_sol} gives us the following modified design formulation
\begin{equation}
\label{eq:filter_proposed}
\begin{aligned}
& \underset{\textbf{w}}{\text{Min}}
& & \textbf{w}^H\textbf{P}\textbf{w} \textrm{ Subject to } \textbf{C}^H \textbf{w}=\textbf{f}
\end{aligned}
\end{equation}
where the constraint matrix $\textbf{C}$ and the response vector $\textbf{f}$ provide the required constraint on the weight vector $\textbf{w}$ so that the resultant design can have the required exact response at the reference frequency. The constraint matrix $\textbf{C}$ in its most basic form corresponds to the real and imaginary parts of the reference frequency vector where we want to constrain the response for this reference frequency vector in the passband of a filter or the look direction of a wideband beamformer to a fixed desired response with its real and imaginary parts contained in the response vector $\textbf{f}$.

For example, consider the design of a lowpass filter. In order to provide correction for the original formulation flaw, we incorporate a constraint for the filter passband response at the reference frequency to be equal to the desired response with unity gain magnitude and linear phase. For a reference frequency  $\omega_r = 0$, $\textbf{c}(\omega)$ in \eqref{eq:filter_vector} changes to
\begin{equation}
\textbf{c}(\omega_r) =  [1, 1, \cdots, 1]^T\;.
\end{equation}
Then, the constraint matrix $\textbf{C}$ just becomes a constraint vector with $\textbf{C} = \textbf{c}(\omega_r)$ with the response vector $\textbf{f}$ containing the desired unity gain as the response of the filter at $\omega_r = 0$ represented by
\begin{equation}
\textbf{c}(\omega_r)^H\textbf{w}= \textbf{f}\;,
\end{equation}
which is simply
\begin{equation}
[1, 1, \cdots, 1] \textbf{w} = 1\;.
\end{equation} 

This constraint will make sure that the designed response of the filter at the reference frequency in the passband is equal to the desired response. As the original formulation will minimize the variation in the response achieved at other frequencies in the passband with respect to the reference frequency, the overall designed response in the passband will be equal to the desired response, thus solving the original formulation problem. 

Note that we can also add other constraints to the formulation of $\textbf{C}$ and $\textbf{f}$ so that more flexible constraints can be imposed on the design. For example, we can add a constraint to make sure the resultant design has an exact zero response at some stopband frequencies.

The solution to \eqref{eq:filter_proposed} can be obtained by the Lagrange multipliers method and it is given by
\begin{equation}
\label{eq:filter_proposed_sol}
\textbf{w}_{opt}=\textbf{P}^{-1} \textbf{C} (\textbf{C}^H \textbf{P}^{-1} \textbf{C})^{-1} \textbf{f}
\end{equation}
For the wideband beamformer design, the modified problem is given by
\begin{equation}
\label{eq:wideband_proposed}
\begin{aligned}
& \underset{\textbf{w}}{\text{Min}}
& & \textbf{w}^H\textbf{G}_{ef}\textbf{w} \textrm{ Subject to } \textbf{C}^H \textbf{w}=\textbf{f}\;,
\end{aligned}
\end{equation}
where $\textbf{C}$ and $\textbf{f}$ again correspond to the constraint matrix and response vector, respectively. For the wideband beamformer case, just like the filter design scenario, this constraint matrix will correspond to the reference frequency steering vector, where $\textbf{C} = \textbf{d}(\Omega_r,\theta_r)$.

By constraining the response of the wideband beamformer at this reference frequency steering vector equal to the desired response $e^{-j \frac{J}{2}\Omega_r}$ as
\begin{equation}
\textbf{d}(\Omega_r,\theta_r)^H \textbf{w} = e^{-j \frac{J}{2}\Omega_r}\;,
\end{equation}
the overall response of the wideband beamformer at the look direction for different frequencies will be equal to the desired response, thus mitigating the initial formulation problem. 
The solution to \eqref{eq:wideband_proposed} is then given by
\begin{equation}
\label{eq:wideband_proposed_sol}
\textbf{w}_{opt}=\textbf{G}_{ef}^{-1} \textbf{C} (\textbf{C}^H \textbf{G}_{ef}^{-1} \textbf{C})^{-1} \textbf{f}
\end{equation}

Note that there are matrix inversion operations in \eqref{eq:filter_proposed_sol} and \eqref{eq:wideband_proposed_sol}, which can be computationally intensive for larger filters and beamformers. However, there are other approaches available in literature e.g. null space based methods to solve \eqref{eq:filter_proposed} and \eqref{eq:wideband_proposed} avoiding the need to compute matrix inversion \cite{liu10g}.

%%%%%%%%%%%%%%%%                            SECTION 4     DESIGN EXAMPLES                              %%%%%%%%%%%%%%%%%%%%%

\section {\textbf{Design Examples}}
\label{sec:design_examples}
In this section, design examples are provided to show the inconsistent performance produced by the original unconstrained eigenfilter design method. The examples are then re-designed through the proposed constrained eigenfilter method to show the improvement.

%%%%%%%%%%%%%%%%%                       SUBSECTION  4A  UNCONSTRAINED EIGENFILTER        %%%%%%%%%%%%%%%%%%%%%%%

\subsection  {\textbf{Unconstrained eigenfilter design}}
\label{sec:unconstrained_eigen}
\begin{figure}[htbp]
   \includegraphics[width=0.48\textwidth]{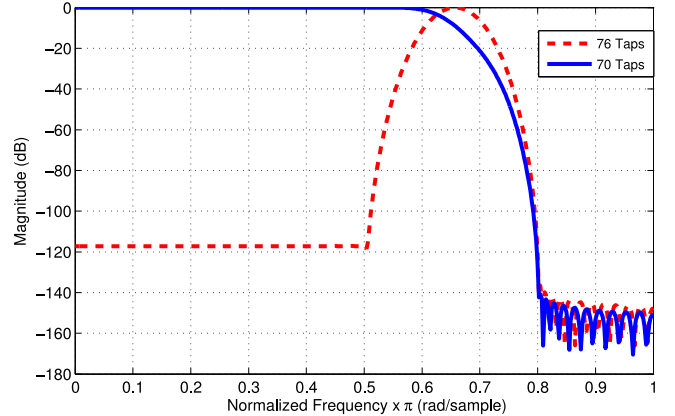}
  \caption{The designed lowpass FIR filters using the original formulation.}
\label{fig:lowpass}
 \end{figure}

First, we consider the lowpass filter design scenario where the whole frequency range from $[0, \pi]$ was discretized into 400 points. The design specifications include the passband from [0, 0.5$\pi$] and stopband from [0.8$\pi$, $\pi$]. A 70-tap filter with trade-off parameter $\alpha$ = 0.97 and reference frequency at 0.35$\pi$ is then designed using the original formulation. The result is shown in Fig. \ref{fig:lowpass} in blue colour (solid curve) with a clearly satisfactory design performance showing a passband to stopband ratio of 140 dB.

In the second case, we just change the number of taps to 76, while keeping all the other specifications the same as the first case. The result is shown in Fig. \ref{fig:lowpass}, highlighted in dashed curve with red colour. We can see that the passband response is out of control, with a flat response of around -118 dB, and the resulting ratio between passband and stopband is just around 19 dB (if ignoring the unacceptable response at the transition band), clearly highlighting the problem with the original formulation.

\begin{figure}[htbp]
   \includegraphics[width=0.49\textwidth]{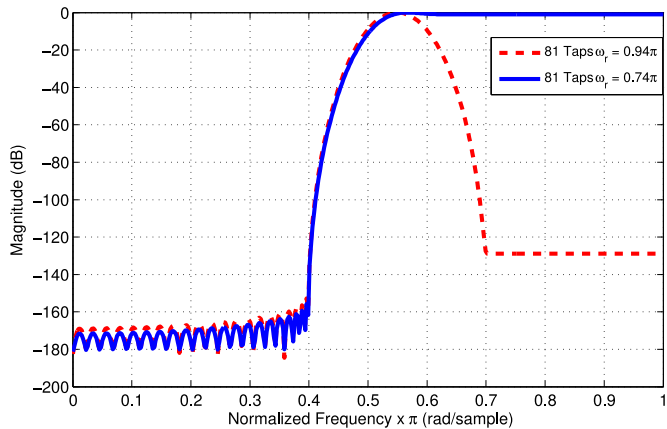}
  \caption{The designed highpass FIR filters using the original formulation.}
\label{fig:highpass}
 \end{figure}

For highpass filters, again two cases are presented. For the first case, we consider an 81-tap filter, where the design specifications include a stopband from [0, 0.4$\pi$] and passband from [0.7$\pi$, $\pi$]. The tradeoff factor $\alpha$ = 0.71 and the reference frequency is set to 0.74$\pi$. The result is depicted in Fig. \ref{fig:highpass} with solid curve and blue colour, where a very satisfactory design performance can be observed with a passband to stopband ratio of 150 dB.

For the second case, we just change the reference frequency to 0.94$\pi$ and the result is shown in  Fig. \ref{fig:highpass} with dashed red colour, which is without any doubt unacceptable, with a passband response at around  -130 dB leaving a passband to stopbad ratio of only 15 dB. The results for lowpass and highpass filter design examples clearly demonstrate the magnitude of the problem at hand for different arbitrary design scenarios. 

Now we extend this observation to the design of bandpass filters to see if the same problem can be observed in those filters as well.
\begin{figure}[htbp]
 \includegraphics[width=0.49\textwidth]{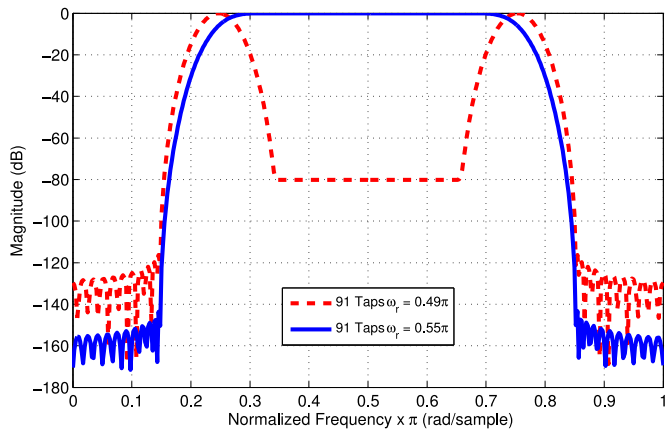}
 \caption{The designed bandpass FIR filters using the original formulation.}
 \label{fig:bandpass}
 \end{figure}

For the bandpass filter design scenario, we again consider two cases for comparison. For the first case, we have 91 taps, where the design specifications include the 1$^{st}$ stopband from [0, 0.15$\pi$], passband from [0.35$\pi$, 0.65$\pi$] and the 2$^{nd}$ stopband from [0.85$\pi$, $\pi$]. The tradeoff factor $\alpha$ = 0.96 and the reference frequency is set to 0.55$\pi$. The satisfactory design result is shown in Fig. \ref{fig:bandpass} with solid curve and blue colour, where a suitable passband to stopband ratio of 145 dB can be observed.

For the second case, we change the reference frequency to 0.49$\pi$, while keeping the remaining specifications similar to the first case and the result is shown with dashed red colour where it can be seen that the flat passband again has dropped to a very low unacceptable magnitude of -80 dB with a passband to stopband ratio of 36 dB, providing further evidence for the kind of inconsistent results caused by the flawed design formulation.
\begin{figure}
\begin{subfigure}[b]{0.245\textwidth}  \includegraphics[width=\linewidth]{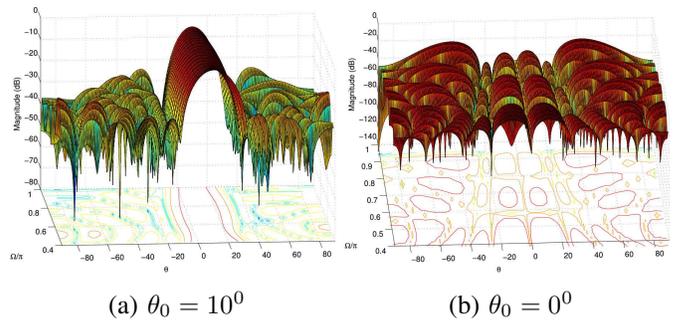}    \caption{$\theta_0=10^0$}    \label{fig:f_1}  \end{subfigure}%
\begin{subfigure}[b]{0.245\textwidth}  \includegraphics[width=\linewidth]{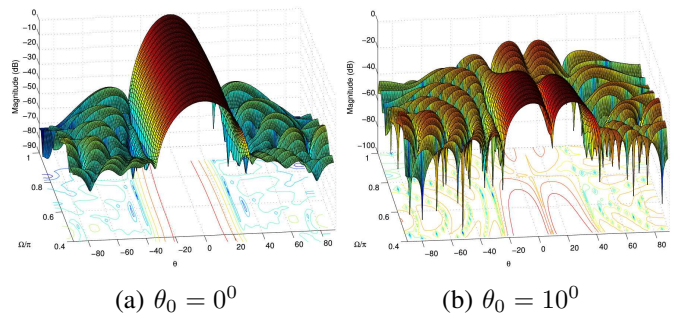}    \caption{$\theta_0= 0^0$}      \label{fig:f_2}   \end{subfigure}
 \caption{The designed wideband beamformer using the original formulation.}
\label{fig:wideband_prob}
 \end{figure}
\begin{figure}
\begin{subfigure}[b]{0.245\textwidth}  \includegraphics[width=\linewidth]{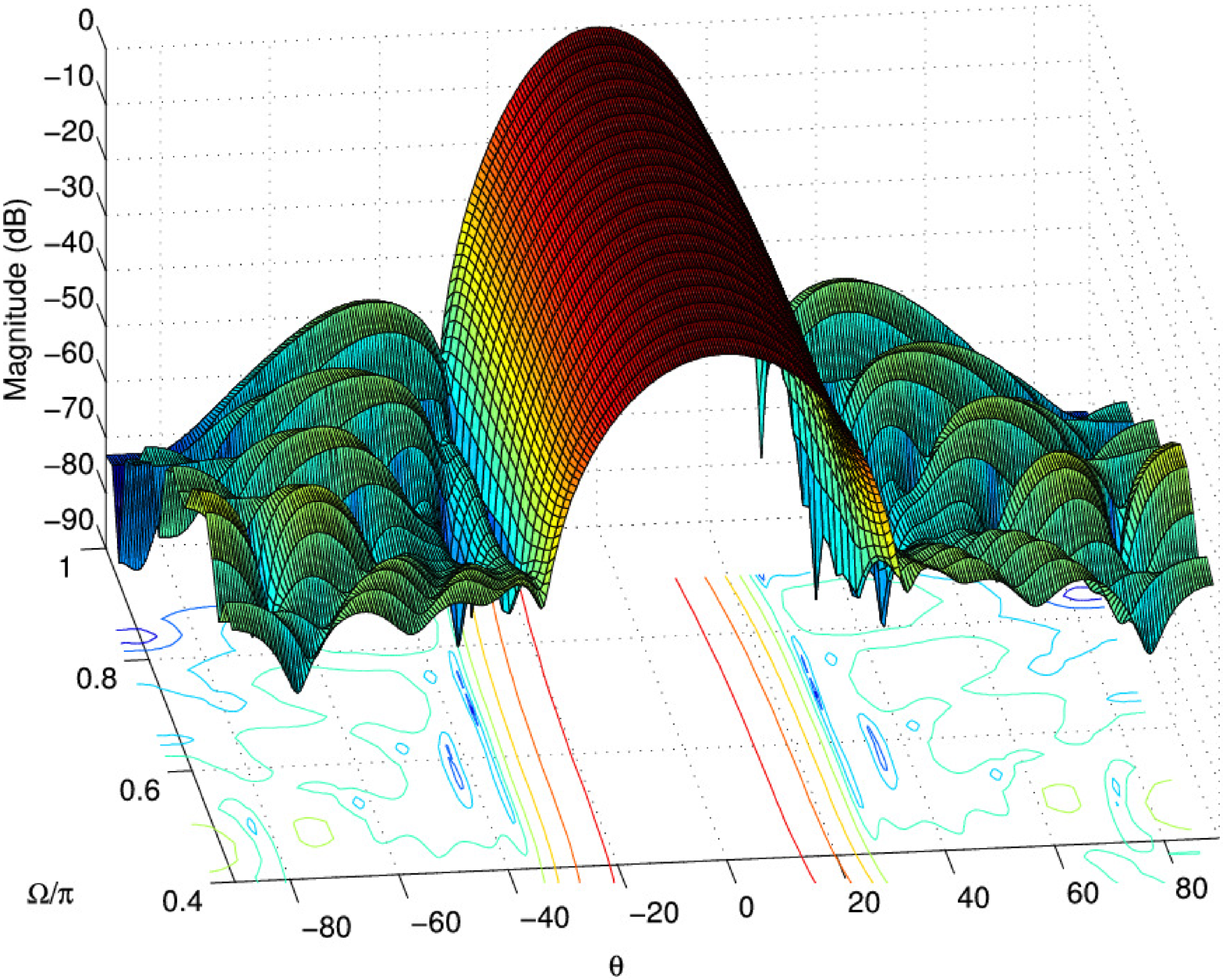}    \caption{$\theta_0=0^0$}    \label{fig:f_1}  \end{subfigure}%
\begin{subfigure}[b]{0.245\textwidth}  \includegraphics[width=\linewidth]{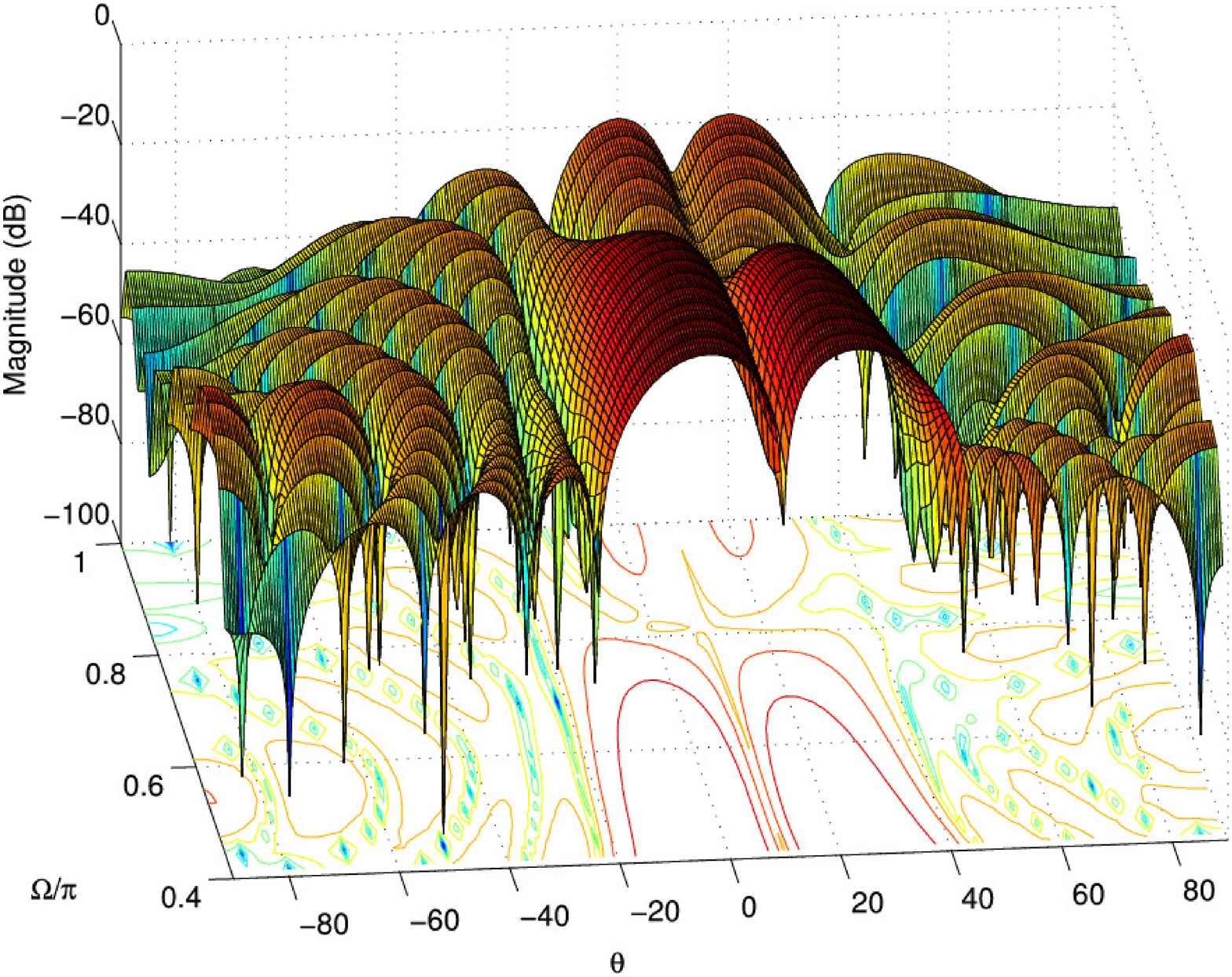}    \caption{$\theta_0=10^0$}      \label{fig:f_2}   \end{subfigure}
 \caption{The designed wideband beamformer using the original formulation.}
\label{fig:wideband_prob1}
 \end{figure}

For the wideband beamformer design, we consider an array with 10 sensors and a TDL length of 10 taps. The look direction is chosen as an off-broadside direction of $\theta_0=10^\circ$ with the desired response equal to $e^{-j5\Omega}$. The considered wideband signal has a frequency range of $\Omega_{pb}=[0.4\pi, \pi]$ with the reference frequency $\Omega_r=0.7\pi$ and $\theta_r=10^\circ$ chosen as the reference point. The weighting function is set to $\alpha=0.6$ at the look direction and 0.4 at the sidelobe region, which runs from from $-90^0$ to $-10^0$ and $30^0$ to $90^0$. The frequency range is discretized into 20 points, while the angle range is divided into 360 points.

\begin{figure*}[htbp]
\begin{tabular}{>{\centering\arraybackslash} m{5.5 cm} >{\centering\arraybackslash} m{5.5 cm} >{\centering\arraybackslash} m{5.5 cm}}
     \toprule
    (a) Lowpass & (b) Highpass  & (c) Bandpass\\
    \hline
     \raisebox{-\totalheight}{\includegraphics[width=0.33\textwidth]{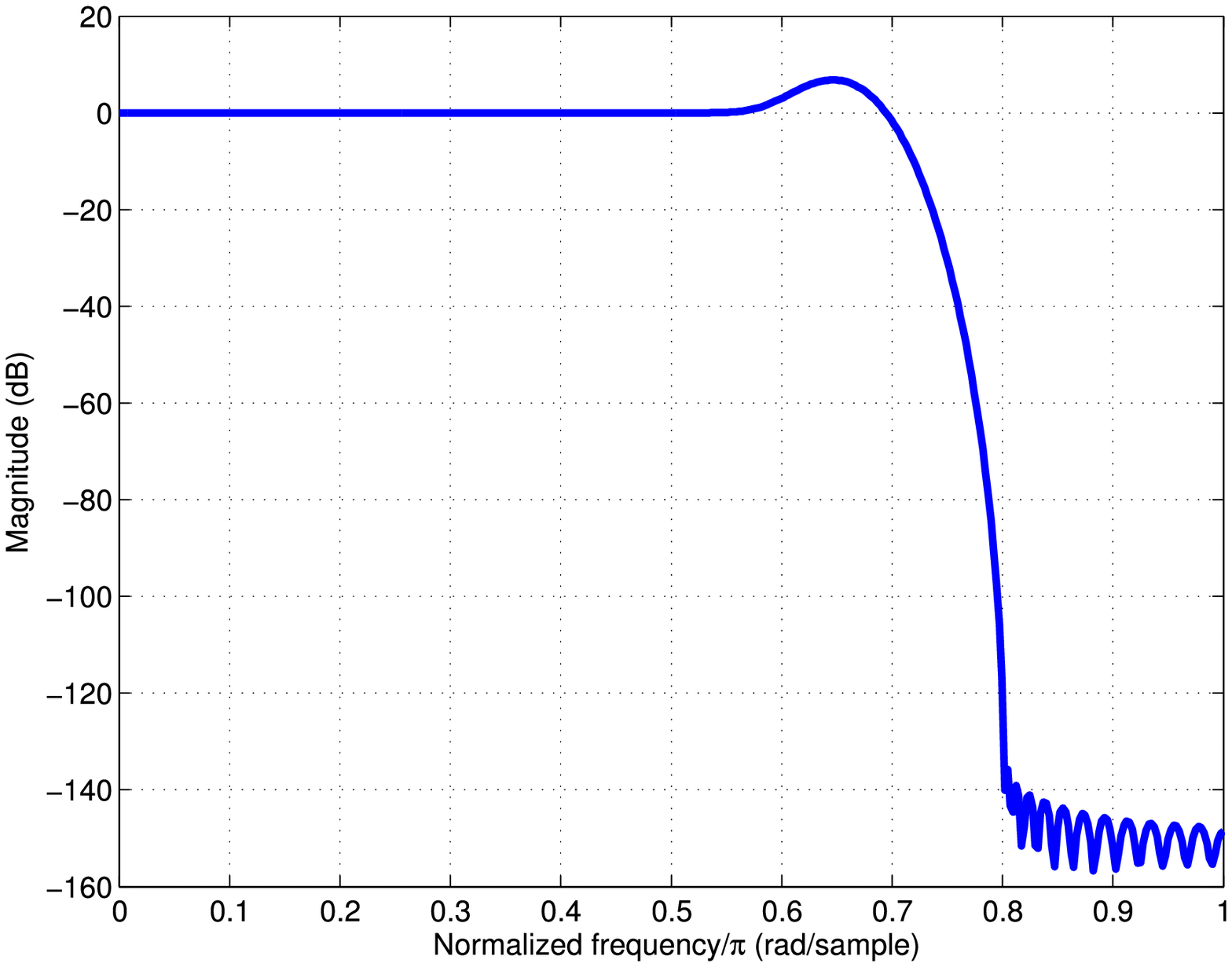}}
      &
     \raisebox{-\totalheight}{\includegraphics[width=0.33\textwidth]{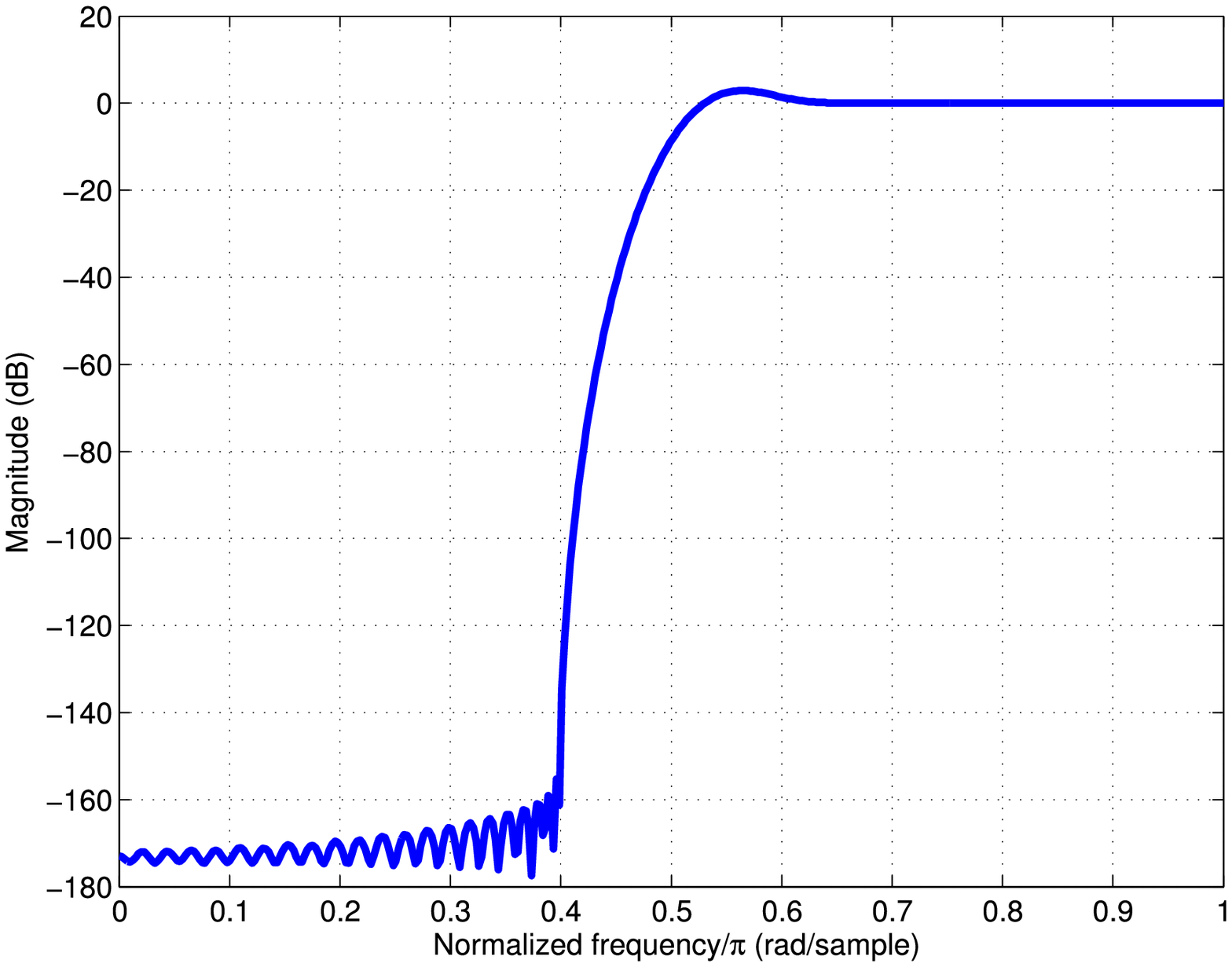}}
      &
     \raisebox{-\totalheight}{\includegraphics[width=0.33\textwidth]{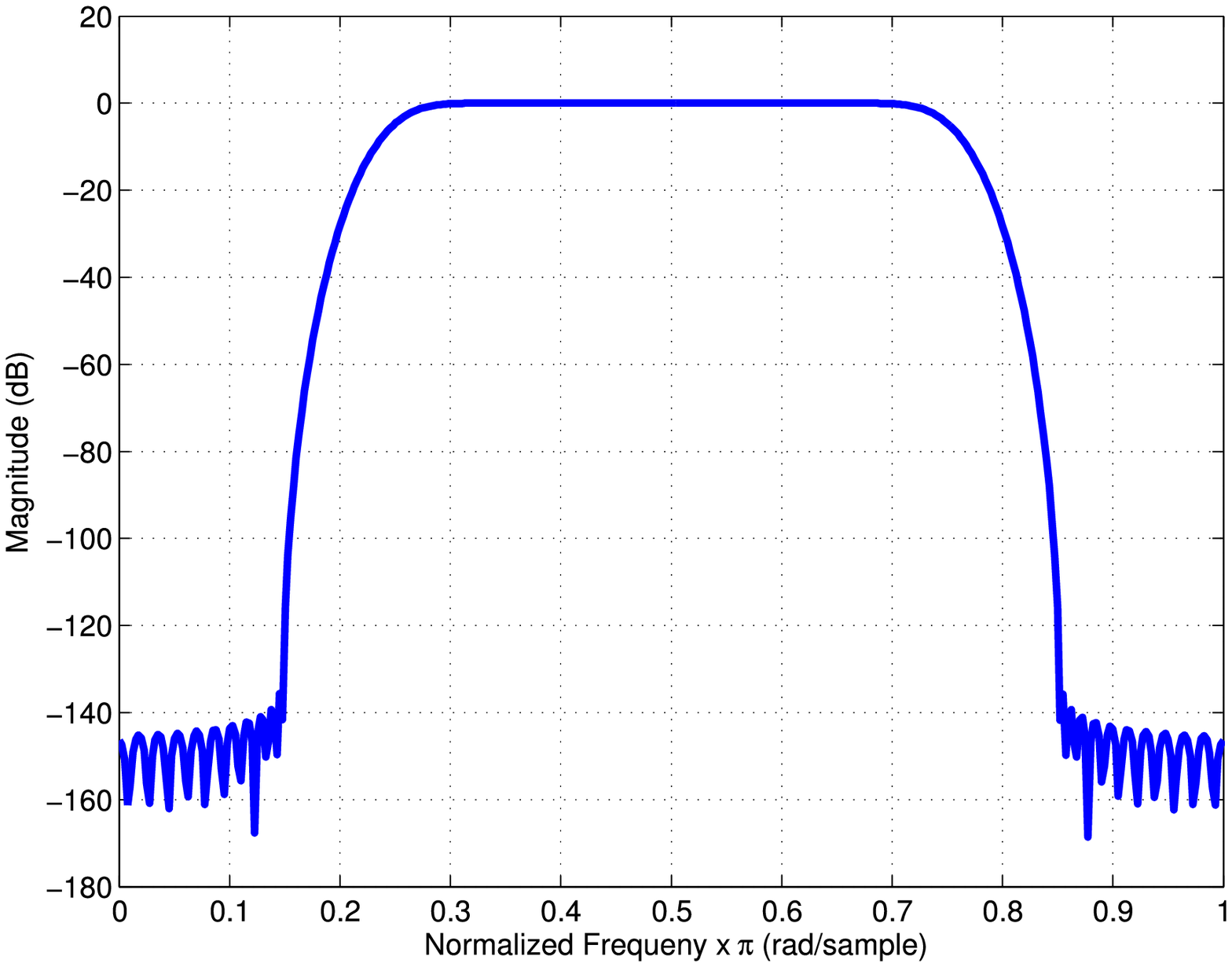}}
     \\ \bottomrule
      \end{tabular}
      \caption{Designed (a) lowpass (b) highpass and (c) bandpass filters using the constrained design.}
      \label{fig:comp}
      \end{figure*}

The result is shown in Fig. \ref{fig:wideband_prob}(a), where a satisfactory design performance is achieved with the look direction to sidelobe ratio around 20 dB.
The same scenario is again tested by changing the look direction to the broadside of $\theta_0=0^0$ with the sidelobe region ranging from $-90^0$ to $-20^0$ and $20^0$ to $90^0$ with the remaining specifications unchanged. The result is shown in Fig. \ref{fig:wideband_prob}(b), where it can be observed that the look direction response plunges to -40 dB with a flat response attained, which is even lower than the sidelobes.

We provide another example for a scenario where we consider an array with 11 sensors and a TDL structure of 10 taps. For the first case, the look direction is chosen as the broadside direction with  $\theta_0=0^\circ$ and the desired response equal to $e^{-j5\Omega}$. For the design specifications we consider a wideband signal having a frequency range of $\Omega_{pb}=[0.4\pi, \pi]$ with the reference frequency $\Omega_r=0.7\pi$ and $\theta_r=10^\circ$ chosen as the reference point. The weighting function is the same as the previous example and the sidelobe region is from $-90^0$ to $-30^0$ and $30^0$ to $90^0$. The result is shown in Fig. \ref{fig:wideband_prob1}(a), where an excellent design response is achieved with a look direction to sidelobe response ratio of 40 dB.  For the second case, we change the look direction to an off-broadside direction of  $\theta_0=10^\circ$ with the sidelobe ranging from $-90^0$ to $-20^0$ and $40^0$ to $90^0$ with the remaining specifications unchanged. The result is shown in Fig. \ref{fig:wideband_prob1}(b), where the look direction response again has no absolute control and achieves flatness around -30 dB with the resulting look direction response even lower than the sidelobes, again demonstrating the presence of this problem in a wide range of design scenarios.

%%%%%%%%%%%%%%%%%                       SUBSECTION  4B  CONSTRAINED EIGENFILTER                         %%%%%%%%%%%%%%%%%%%%%%%

\subsection{\textbf{Constrained eigenfilter design}}
\label{sec:constrained_eigen}
We now apply the constrained eigenfilter formulation in \eqref{eq:filter_proposed} to design the lowpass, highpass and bandpass filters presented using unconstrained design formulation. The new results are presented in Fig. \ref{fig:comp}(a), (b) and (c). Although there is still a noticeable bump in the transition band for the design results in Fig. \ref{fig:comp}(a) and (b) for lowpass and highpass, respectively, the overall response has improved significantly compared to the results in Figs. \ref{fig:lowpass} and \ref{fig:highpass}. The bandpass filter designed with the new formulation in Fig. \ref{fig:comp}(c) achieves a very satisfactory response compared to the result in Fig. \ref{fig:bandpass}.
\begin{figure}[htbp]
\centering
 \includegraphics[width=0.49\textwidth]{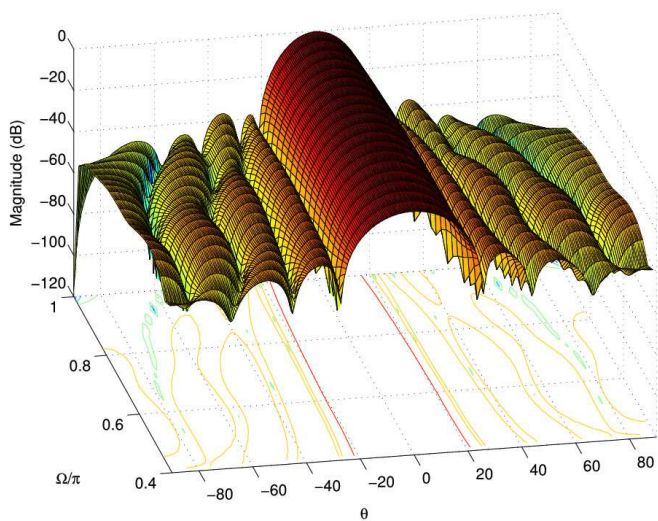}
 \caption{The designed wideband beamformer with $\theta_0= 0^0$.}
\label{fig:wideband_proposed_result}
\end{figure}
\begin{figure}[htbp]
\centering
 \includegraphics[width=0.49\textwidth]{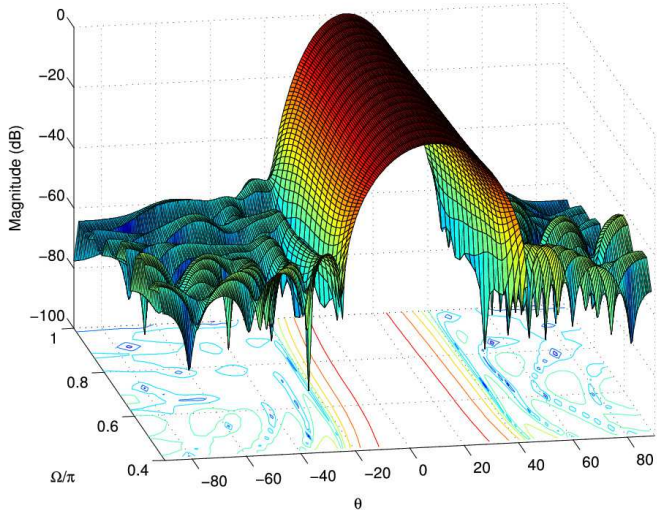}
 \caption{The designed wideband beamformer with $\theta_0= 10^0$.}
\label{fig:wideband_proposed_result1}
\end{figure}

For the beamformer design presented in Figs. \ref{fig:wideband_prob}(b) and \ref{fig:wideband_prob1}(b), we re-design them using the constrained formulation in \eqref{eq:wideband_proposed} and the result is provided in Figs. \ref{fig:wideband_proposed_result} and \ref{fig:wideband_proposed_result1}, where the look direction response has improved significantly with a decent look direction to sidelobe ratio achieved as per the desired specifications.

We have tried various designs for different types of filters and wideband beamformers with varying design specifications and the proposed method has been found to perform consistently well in different scenarios.

%%%%%%%%%%%%%%%%%%%                           SECTION 5 CONCLUSION                         %%%%%%%%%%%%%%%%%%

\section{\textbf{Conclusion}}
\label{sec:conclusion}
The classic eigenfilter approach has been revisited and critically analyzed, where a formulation problem is highlighted in the passband/look direction part of the cost function which leads to an inconsistent design performance. A solution was then proposed by adding a linear constraint, explicitly setting the designed passband response at the reference frequency point to the desired one. Results have been provided for different design scenarios based on FIR filter and wideband beamformer design to demonstrate the crucial issue of the original formulation and the satisfactory performance by the proposed one.

%%%%%%%%%%%%%%%%%%%%                                BIBLIOGRAPHY                                          %%%%%%%%%%%%%%%%%%%%%%

\bibliography{mybib}

% that's all folks
\end{document}